\documentclass[conference]{IEEEtran}
\usepackage{cite}
\usepackage{graphicx}
\usepackage{amsmath}
\usepackage{url}

\begin{document}

\title{Bridging Cultural and Digital Divides: A Low-Latency JackTrip Framework for Equitable Music Education in the Global South}

\author{
  \IEEEauthorblockN{Tiange Zhou}
  \IEEEauthorblockA{
    School of Future Design\\
    Beijing Normal University\\
   Zhuhai, China\\
    tiangezhoumusic@gmail.com
  }
  \and
  \IEEEauthorblockN{Marco Bidin}
  \IEEEauthorblockA{
    Department of Instrument Engineering\\
    Xinghai Conservatory of Music\\
    Guangzhou, China\\
    artdir.alea@yahoo.it
  }
}

\maketitle

\begin{abstract}
The rapid expansion of digital technologies has transformed educational landscapes worldwide, yet significant infrastructural and cultural challenges persist in the Global South. This paper introduces a low-latency JackTrip framework designed to bridge both the cultural and digital divides in music education. By leveraging an open‐source, UDP‐based audio streaming protocol originally developed at Stanford’s CCRMA, the framework is tailored to address technical constraints—such as intermittent connectivity, limited bandwidth, and high latency—that characterize many rural and underserved regions. The study systematically compares the performance of JackTrip with conventional platforms like Zoom, demonstrating that JackTrip achieves sub-30~ms latency under simulated low-resource conditions while preserving the intricate audio details essential for non-Western musical traditions. Spectral analysis confirms that JackTrip’s superior handling of microtonal scales, complex rhythms, and harmonic textures provides a culturally authentic medium for real-time ensemble performance and music education. These findings underscore the transformative potential of decentralized, edge-computing solutions in empowering educators and musicians across the Global South, promoting both technological equity and cultural preservation.
\end{abstract}

\IEEEpeerreviewmaketitle

\section{Introduction}
Music education has traditionally been a cornerstone of cultural transmission, particularly in those regions of the world where oral traditions are dominant. In countries across the Global South, including sub-Saharan Africa, South Asia, and Latin America, music is far more than an artistic endeavor; it is an integral component of identity, history, and spirituality. In West Africa, for example, Griot drummers inscribe historical events in the form of intricate poly-rhythmic patterns, while in India, classical musicians perform in micro-tonal intervals to express deep emotional and spiritual nuances\cite{ref1}. These musical systems cannot be separated from the cultural context in which they arise, and performance will typically involve both musical and non-musical communication, such as oral storytelling and spiritual expression.

However, these rich musical heritages are subject to existential threats, largely precipitated by the exigencies of modern technologies and infrastructural deficits. According to the Science Direct, in Sub-Saharan Africa (SSA), an estimated 32 percent of primary schools and almost half of secondary schools still lacked access to electricity for promising internet connection.\cite{ref2}. These infrastructural deficiencies severely hamper the utilization of modern educational technologies and digital platforms, and prevent them from promoting global participation in the arts and advancing the preservation of culture.

Widely available digital platforms like Zoom, Skype, and Google Meet, though capable of bridging geographical distances between individuals across the world, are designed with speech and general communication rather than the specific needs of music performance in mind. These platforms are based on audio standards that privilege 12-tone equal temperament (12-TET), a tuning system prevalent in Western music but ill-equipped for the micro-tonal, non-Western scales of much Global South musical practice. Additionally, these platforms have high latencies and poor handling of rhythmically complex music, making them inappropriate for real-time ensemble music education in low-resource settings. 

In response to such severe challenges, this study explores JackTrip, a low-latency UDP-based audio streaming open-source software developed at Stanford's Center for Computer Research in Music and Acoustics (CCRMA)\cite{ref3}. JackTrip is explicitly developed for real-time networked music performance even under conditions of high latency and low bandwidth and is therefore a prime candidate to facilitate remote music education in the Global South. This study evaluates JackTrip from both a technical and a cultural perspective, exploring its potential for supporting the performance of non-Western music, cross-cultural collaboration, and the preservation and dissemination of cultural heritage in the resource-poor Global South.

In particular, this study points out the potential of JackTrip for enabling music ensemble performance in the Global South, where Internet connectivity problems often make traditional online music education software unusable. By utilizing inexpensive edge devices such as the Raspberry Pi, JackTrip offloads processing from cloud-based servers, thereby reducing latency as well as expense. The ability to perform in real-time, even in rural and underserved areas, has significant implications not only for music education but also for cultural exchange and economic development in these communities.

\section{Related Works}

\subsection{The Significance of Music Education in the Global South}
Music education in the Global South is not only a method of artistic instruction but a critical component in the maintenance of cultural identity and the transmission of knowledge. In most of the world, particularly sub-Saharan Africa, South Asia, and parts of Latin America, music is part of communal life, often reflecting the historical, religious, and social life of these communities. For instance, the West African Griot tradition is deeply embedded in the social life, with griots performing both the role of musicians and historians, encoding and transmitting the history and knowledge of their people in music\cite{ref4}. Music education in these communities is often an oral tradition, transmitted from one generation to the other through performance rather than classroom instruction.

This educational music approach is obviously different from that of the Western education system, where formal, written notation systems and rigorous pedagogical methods dominate. Indian classical music education is based on intricate systems of ragas (melodic modes) and talas (rhythmic cycles) that are transmitted orally and learned through imitation and apprenticeship\cite{ref5}. These musical traditions do not merely represent artistic genres but worldviews, spiritual systems, and understandings of the human condition. This cultural specificity underscores the importance of maintaining these musical traditions through education that is faithful to their indigenous forms and structures. However, music education in the Global South is faced with acute challenges in the form of widespread infrastructural deficiencies. Further, traditional pedagogical models that assume face-to-face mentorship and group learning are strained in locations where access to such face-to-face instruction is limited by geographical or economic distance.

One of the most pressing issues, however, is the struggle between maintaining indigenous musical traditions and the growing ubiquity of Western musical systems. As digital technologies become more pervasive, Western music education models—based on standardized notation systems and pitch sets like 12-TET—have come to disproportionately dominate the globe. This kind of musical imperialism, which we can refer to as sonic imperialism, leads to the marginalization or distortion of non-Western musical practices. Specifically, tools like Zoom and Skype, the de facto standard for online learning, are optimized for speech and general audio communication but are not accommodative of the diversity of musical forms within the Global South. By normalizing the tuning system to Western 12-TET, these platforms compress the tonal and rhythmic complexities central to non-Western music traditions, such as micro-tonal scales found in Indian classical music or the intricate poly-rhythms found in African drumming. This poses a significant problem: the music and learning experiences of the Global South are distorted and, in some cases, erased.

The challenge, therefore, is twofold: how to preserve the rich and diverse musical heritage of the Global South in a time of rapidly advancing technology, and how to develop systems of music education that honor and nurture these heritage systems. The increasing digital divide only makes this issue more urgent, as the vast majority of the Global South is excluded from the technological advancement that would otherwise make their inclusion in a global system of music education feasible.

\subsection{Challenges in Adopting Technology-Based Music Education in the Global South}
The challenges of the Global South in embracing digital technologies for music education extend beyond issues of infrastructure and cultural preservation. While much attention has been given to issues of access to hardware and the internet, of equal importance are the socio-technical barriers that influence the adoption of new technologies. Among the most significant barriers is the lack of locally appropriate solutions that respond to the unique technological, economic, and cultural conditions of the Global South. For example, commercial solutions such as Zoom, Google Meet, and Skype are often not suited to the specific needs of music education. These tools were not designed with the complexities of real-time music performance in mind. They are optimized for general voice communication and often use compression algorithms that degrade the quality of audio, as a consequence, flattening out the nuances so essential to musical expression.

Along with the technical concerns are also matters of digital literacy. Throughout much of the Global South, there are significant gaps in the technical knowledge required to optimize the use of online music education technologies. Even where access to affordable technology is available, it is often underused or misused due to a lack of training. Added to this is the fact that many of those who work in teaching positions in these countries may simply not have had the same pedagogical training or access to resources as instructors in more developed countries, and therefore there is a lack of both the knowledge of, as well as the practical application of, new digital music education technologies.

Furthermore, there are economic issues that make extensive use of such tools impossible. Educational budgets in the majority of low- and middle-income nations are stretched thin, and music and arts education are always the last ones to be allocated any funds. While the cost of digital tools such as smartphones and computers has decreased, the lack of availability of constant electricity, affordable internet, and the prohibitive cost of cloud services are all significant barriers to their adoption.

The other key consideration is the issue of access to energy. Power outages in rural and underserved areas in the Global South are frequent, and access to a stable supply of electricity is limited. The International Energy Agency (IEA) estimates that nearly 600 million people in sub-Saharan Africa lack any access to electricity, so any attempt to introduce technology-based learning to such communities must consider energy limitations\cite{ref6}. In the absence of access to stable sources of power, devices like computers, tablets, and servers that are required to support online educational platforms are not viable. Any solution to this dilemma, therefore, must be designed with power efficiency in mind, which makes low-power devices like Raspberry Pi and solar-powered initiatives particularly valuable.

\section{JackTrip as a Solution for Low-Latency Music Performance under Resource-Constrained Environments}
JackTrip, unlike traditional cloud-based applications over TCP/IP (Transmission Control Protocol/Internet Protocol) and centralized server-based processing, JackTrip’s decentralized architecture takes advantage of edge computing to minimize latency and reliance on high-bandwidth internet connectivity. This makes it highly suitable for networked music performance in low-infrastructure and high-latency environments,which will support real-time audio transmission with sub-50~ms latency, an essential requirement for enabling ensemble music performance even in high-latency network environments\cite{ref7}.

Comparing to Zoom, which require a central server to process data, JackTrip allows processing to occur on local edge devices. This reduces the burden on central servers and allows the use of low-cost devices, such as the Raspberry Pi, to receive and send audio signals. Raspberry Pi devices are inexpensive, power-efficient, and widely available, and thus offer a great solution for low-resource settings. By leveraging such low-cost hardware, JackTrip is capable of providing an inexpensive and scalable solution for music education that does not rely on expensive infrastructure or cloud services.

One of the most important benefits of JackTrip, especially for the Global South, is that it can preserve the distinctive features of non-Western musical traditions. JackTrip's open-source nature allows it to be customized to a great extent, including adapting its software to process alternate tuning systems, rhythms, and musical structures common in the Global South. Indian classical music, for example, relies on microtonal scales and complex rhythmic cycles and requires a platform that can handle subtleties of pitch and timing that platforms like Zoom are not equipped to process. JackTrip, however, allows users to implement these features by using software patches such as Pure Data (Pd), which can be adapted to include micro-tonal scales or rhythm-sensitive buffer algorithms~\cite{ref8}.

Apart from this, JackTrip has already demonstrated its cultural relevance in a series of cross-cultural performances. For instance, the "Changing Tides"concert organized by UCSD united Korean folk singers and a Mexican flute player, using JackTrip to facilitate real-time performance among geographically distant musicians. The collaboration demonstrated how JackTrip could not only facilitate ensemble performance but also serve as a cultural bridge, allowing musical traditions of varying sorts to be heard and accessed by global audiences. This kind of cultural exchange is crucial in promoting mutual respect and understanding among cultures, especially during a time when globalization tends to lead to the dominance of a few Western musical paradigms~\cite{ref9}.
\begin{figure}[h!]
  \centering
  \includegraphics[width=0.8\linewidth]{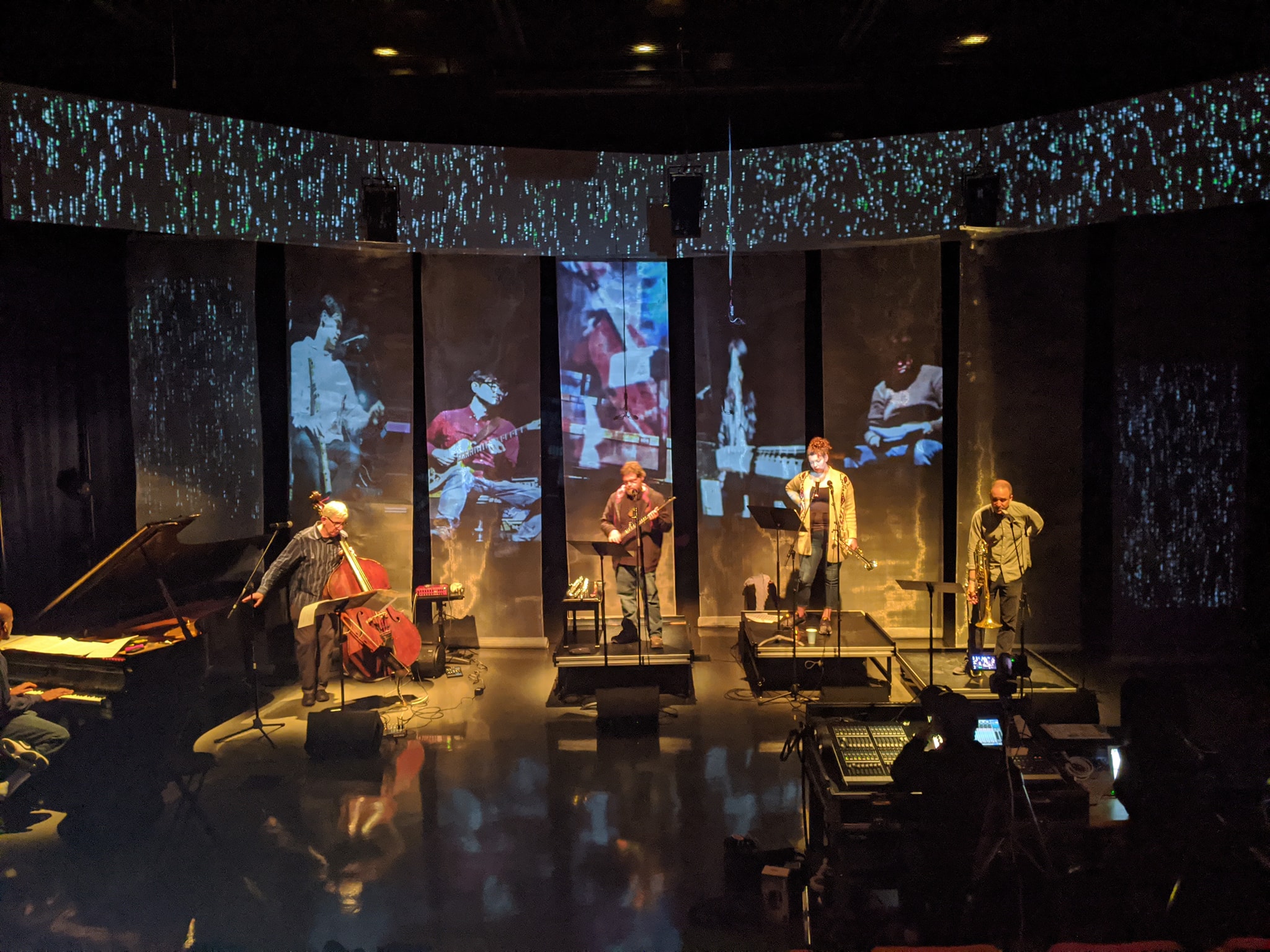}
  \caption{"Changing Tides" Telematic music concert March 2020.}
  \label{fig:fig1}
\end{figure}

Despite its potential, however, JackTrip has not been extensively used in low-resource settings. Much of the previous research on networked music performance has presumed high-bandwidth environments in which latency is less problematic (e.g., fiber-optic networks). This paper tries to address this
gap by evaluating JackTrip’s potential to support real-time
music education in the Global South and analyzing how its
technical and cultural affordances make it an exemplary tool
for Non-western centric music education in these regions.

\section{Methodology}
The study utilized a controlled experimental approach to assess and compare the effectiveness of JackTrip and Zoom audio streaming platforms under network conditions typical of rural regions in the Global South. Conditions simulated included limited bandwidth, high latency, and intermittent connectivity. Two computers were positioned approximately 1,000 kilometers apart to represent realistic geographic separation between remote educational institutions. Network constraints characteristic of typical 3G and 4G cellular connections were simulated using Network Link Conditioner software (Mac OS) and Clumsy software (Windows OS), setting parameters for bandwidth limits (uplink: 5--10~Mbps, downlink: 10--50~Mbps) and packet loss (0--3\%)\cite{ref10}.

The total latency ($T_{\text{total}}$) for audio communication was mathematically determined as the sum of three primary components:
\begin{equation}
T_{\text{total}} = T_d + T_p + T_{proc}
\end{equation}
\begin{itemize}
    \item \textbf{Transmission Delay ($T_d$):} Calculated by dividing packet size by bandwidth. Assuming an audio streaming packet size of 512 bytes (4096 bits) and minimum bandwidth (5 Mbps), the transmission delay was computed as:
    \[
    T_d = \frac{4096\ \text{bits}}{5\times10^6\ \text{bps}} \approx 0.819\ \text{ms}
    \]
    \item \textbf{Propagation Delay ($T_p$):} Determined by dividing distance by the propagation speed (approximately $2 \times 10^8$ m/s for fiber optics). For a 1,000~km distance:
    \[
    T_p = \frac{1\,000\,000\ \text{m}}{2\times10^8\ \text{m/s}} = 5\ \text{ms}
    \]
    \item \textbf{Processing Delay ($T_{proc}$):} Measured empirically via repeated ping tests between the computers, with JackTrip delays ranging from 15--20~ms and Zoom delays ranging from 130--140~ms.
\end{itemize}

Audio streams from both platforms were simultaneously recorded under identical network conditions. Three categories of audio samples were captured:
\begin{enumerate}
    \item Short-duration transient sounds (under 1 second)
    \item Medium-length sustained notes (3--5 seconds)
    \item Extended-duration micro-tonal scales performed with a panpipes from Peru
\end{enumerate}
Detailed spectral analyses were conducted using Audacity software to assess amplitude, frequency, and harmonic content.

\section{Results}
\subsection{Latency Analysis}
Latency calculations highlighted significant performance differences between Zoom and JackTrip:
\begin{itemize}
    \item \textbf{Zoom Total Latency:} 
    \[
    T_{\text{total (Zoom)}} = 0.819\ \text{ms} + 5\ \text{ms} + 135\ \text{ms} \approx 141\ \text{ms}
    \]
    \item \textbf{JackTrip Total Latency:}
    \[
    T_{\text{total (JackTrip)}} = 0.819\ \text{ms} + 5\ \text{ms} + 20\ \text{ms} \approx 26\ \text{ms}
    \]
\end{itemize}
JackTrip exhibited significantly lower latency (approximately 26~ms), which is suitable for real-time ensemble performances. In contrast, Zoom’s higher latency (approximately 141~ms) negatively impacted real-time musical synchronization.

\subsection{Audio Information Analysis}
Spectral analysis clearly identified substantial differences between Zoom and JackTrip:
\begin{itemize}
    \item \textbf{Zoom:} Significant audio compression reduced clarity, specifically above 3~kHz, negatively impacting the accuracy and fidelity of musical details.
    \item \textbf{JackTrip:} Preserved full-spectrum audio details (20~Hz to 20~kHz), maintaining microtonal accuracy, rhythmic complexity, and harmonic richness.
\end{itemize}
\begin{figure}[h!]
    \centering
    \includegraphics[width=1\linewidth]{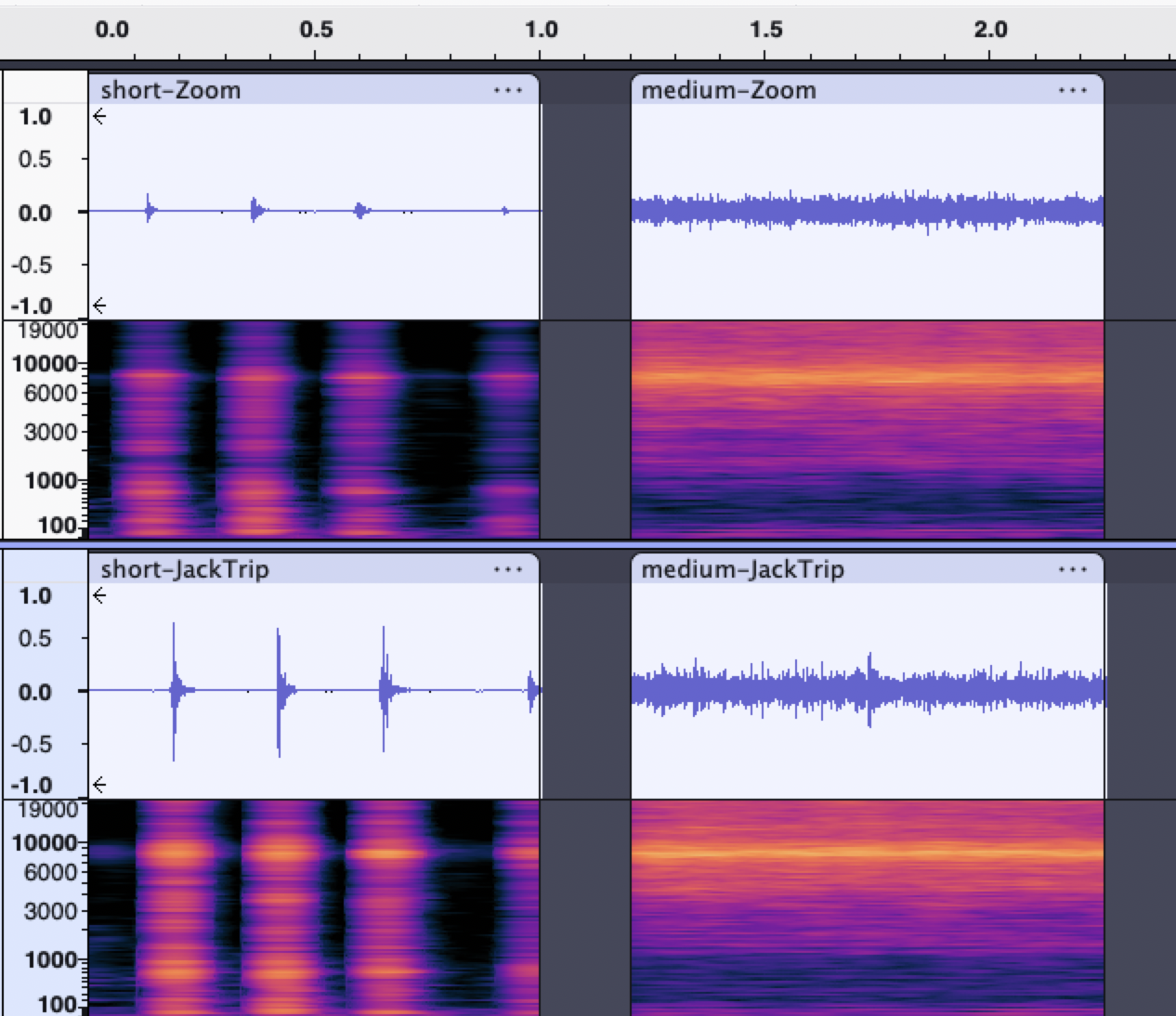}
    \caption{Comparative Analysis of short and medium duration sound through Zoom and JackTrip}
    \label{structure}
\end{figure}
\begin{figure}[h!]
    \centering
    \includegraphics[width=1\linewidth]{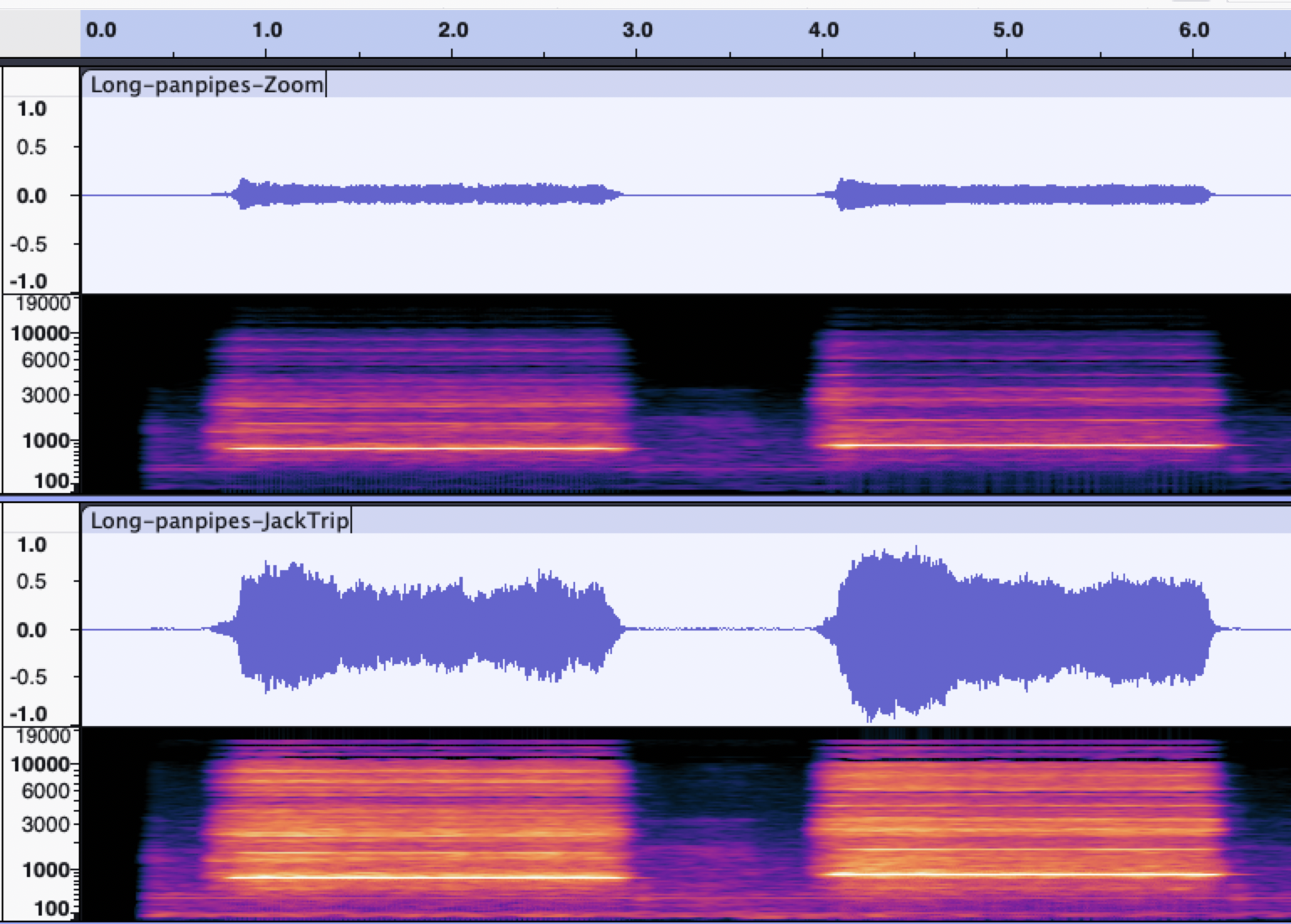}
    \caption{Comparative Analysis of long duration panpipes sound through Zoom and JackTrip}
    \label{structure}
\end{figure}

\subsection{Audio Quality and Cultural Fidelity}
JackTrip recordings effectively retained microtonal accuracy, rhythmic complexity, and harmonic authenticity necessary for the accurate representation of non-Western musical traditions. In contrast, Zoom's compressed audio diminished these essential musical details, thereby undermining cultural authenticity. These outcomes emphasize JackTrip’s technical and cultural relevance for supporting diverse and nuanced musical traditions within resource-limited Global South contexts.

\section{Discussion}
\subsection{Technological Performance and Its Applicability to Music Education in the Global South}
This research presents the key audio information differences between JackTrip and Zoom related to latency, packet loss, and rhythmic and tonal accuracy—factors that are especially critical in network conditions similar to those encountered in remote Global South areas. The low-latency, real-time ensemble performance offered by JackTrip provides a potentially game-changing solution for e-learning in music education in the Global South. The average audio latency in Zoom, around 115--141~ms, is beyond the threshold for real-time human musical connection, while JackTrip’s average latency of approximately 25~ms allows musicians and educators to collaborate in real time, even from geographically distant locations. Furthermore, JackTrip’s capacity to deliver 1536~kbps audio quality—compared to Zoom's 192~kbps—has been verified through detailed spectral analysis in Audacity.

\subsection{Cultural Relevance: Maintaining Non-Western Musical Traditions}
The cultural implications of JackTrip extend beyond its technical capabilities. Inadequate audio transmission can lead to reductive interpretations and erosion of cultural integrity. By enabling high-quality, low-latency audio communication, JackTrip allows musicians and educators in the Global South to maintain local musical practices, including complex rhythms, microtonal nuances, and improvisational traditions, rather than having their traditions oversimplified by Western-centric music theory, such as 12-TET.

\subsection{Empowering the Global South: Education and Societal Transformation}
The power of virtual music education as an innovative factor for the Global South lies in its potential for fundamental social, economic, and cultural transformation. In regions where traditional voice-based music education is rare due to infrastructural and economic constraints, JackTrip offers a new frontier for collaboration among students and teachers regardless of geographic location or financial limitations. By fostering global educational networks, JackTrip can serve as a catalyst for social change, promoting cultural pluralism and the democratization of music education.

\subsection{Challenges and Future Directions}
While this research provides a strong foundation for the potential of JackTrip in enhancing music education in the Global South, challenges remain. The implementation of low-cost, energy-efficient hardware such as the Raspberry Pi is subject to constraints like electricity availability, network connectivity, and maintenance skills. Additionally, integrating digital technologies must be done in a way that supplements rather than eliminates traditional community-based learning practices. Future studies should involve pilot tests in local communities, collaborative development of culturally relevant pedagogical methods, and longitudinal research on the impact of digital technology on traditional music education practices.

\section{Conclusion}
This study confirms that JackTrip, an open-source low-latency audio streaming software, is an effective tool for networked music education in the Global South. Through a comparative evaluation of JackTrip and Zoom under emulated network conditions, the research reveals that JackTrip offers significantly lower latency, better audio fidelity, and enhanced cultural authenticity. Consequently, JackTrip can facilitate real-time ensemble performance even in resource-poor settings where conventional platforms like Zoom fail to provide the necessary capabilities.

JackTrip's high packet loss tolerance, latency reduction, and support for complex rhythms and microtonal tunings make it ideally suited for the nuanced requirements of non-Western musical traditions. Its open-source, flexible architecture allows local educators and musicians in the Global South to adapt the platform to their specific cultural needs. Ultimately, JackTrip not only represents a technical innovation but also a cultural one, bridging digital divides and empowering marginalized communities through equitable access to global music education.

\end{document}